\documentclass[acmsmall]{acmart}
\AtBeginDocument{%
  }

\setcopyright{acmlicensed}
\copyrightyear{2025}
\acmYear{2025}
\acmDOI{XXXXXXX.XXXXXXX}
\acmConference[Nature Communication]{Make sure to enter the correct
  conference title from your rights confirmation email}{June 03--05,
  2025}{Woodstock, NY}
\acmISBN{978-1-4503-XXXX-X/2025/06}

\begin{document}

\title{Enhancing Interpretability in Software Change Management with Chain-of-Thought Reasoning}

\author{Yongqian Sun}
\authornote{TKL-SEHCI is short for Tianjin Key Laboratory of Software Experience and Human Computer Interaction.}
\affiliation{
    \institution{Nankai University \& TKL-SEHCI} \city{Tianjin} \
    \country{China} 
}

\author{Weihua Kuang}
\affiliation{
    \institution{Nankai University} 
    \city{Tianjin} 
    \country{China} 
}

\author{Chao Shen} 
\affiliation{ 
    \institution{Nankai University} 
    \city{Tianjin} 
    \country{China} 
}  

\author{Xidao Wen}
\affiliation{
    \institution{Tsinghua University} 
    \city{Beijing} 
    \country{China} 
} 

\author{Tinghua Zheng}
\affiliation{%
\institution{
    Nankai University} 
    \city{Tianjin} 
    \country{China} 
} 

\author{Heng Liu}
\affiliation{
    \institution{CHINA TIANCHEN ENGINEERING CORPORATION LTD.} \city{Tianjin} 
    \country{China} 
} 

\author{Shenglin Zhang}
\authornote{Shenglin Zhang is the corresponding author. Email: zhangsl@nankai.edu.cn. HL-IT stands for Haihe Laboratory of Information Technology Application Innovation.}
\affiliation{
    \institution{Nankai University \& HL-IT} 
    \city{Tianjin} 
    \country{China} 
}

\author{Bo Wu}
\affiliation{ 
    \institution{Tencent Technologies} 
    \city{Beijing} 
    \country{China} 
}

\author{Dan Pei}
\authornote{BNRist is Beijing National Research Center for Information Science and Technology.}
\affiliation{
    \institution{Tsinghua University \& BNRist} 
    \city{Beijing} 
    \country{China} 
}

\renewcommand{\shortauthors}{Kuang, Shen et al.}

\begin{abstract}
    In modern large-scale online services, frequent software changes have become the norm, but they are also a major source of risk that can lead to service failures and significant economic losses. Therefore, efficient and precise management of software changes is crucial.

    Current software change management processes — including Error Change Detection (ECD), Fault Triage (FT), and Root Cause and Corrective Action (RCCA) — are typically isolated and manual, lacking a unified, automated framework. Although large language models (LLMs) have shown great potential for automation tasks, their “black-box” nature results in opaque decision-making processes, which severely limits their application in operations and maintenance (O\&M) scenarios where high reliability and interpretability are required.

    To address these challenges, we propose \textbf{SCELM} (Software Change Evaluation and Lifecycle Management), an end-to-end automated software change management framework integrated with large language models. SCELM leverages Chain-of-Thought (CoT) reasoning to simulate expert analytical processes and uses reinforcement learning (specifically the GRPO and KTO algorithms) to optimize the model’s reasoning accuracy and interpretability. In addition, the framework employs Retrieval-Augmented Generation (RAG) to dynamically draw knowledge from historical cases and implements continuous learning through a lightweight KTO-based feedback mechanism.

    We conducted extensive experimental evaluations on three real-world industrial datasets. The results show that our approach significantly outperforms existing baselines across the three core tasks of ECD, FT, and RCCA. Our proposed core method, SCoT, not only improves final decision accuracy but also demonstrates, via our Chain-of-Thought evaluation metric CoTScore, that its generated reasoning paths align closely with expert reasoning. Ablation studies and continual learning experiments further verify the effectiveness and lifelong learning capabilities of key components in the framework — including RAG, GRPO, KTO — and the integration of downstream tasks.

    This work provides the first complete solution for an automated, interpretable, and continuously optimizable software change management system. The SCELM framework will greatly advance the practical application of LLMs in the AIOps domain. Moreover, we release the first CoT dataset for change scenarios to facilitate future research in the community.
\end{abstract}


\begin{teaserfigure}
  \includegraphics[width=\textwidth]{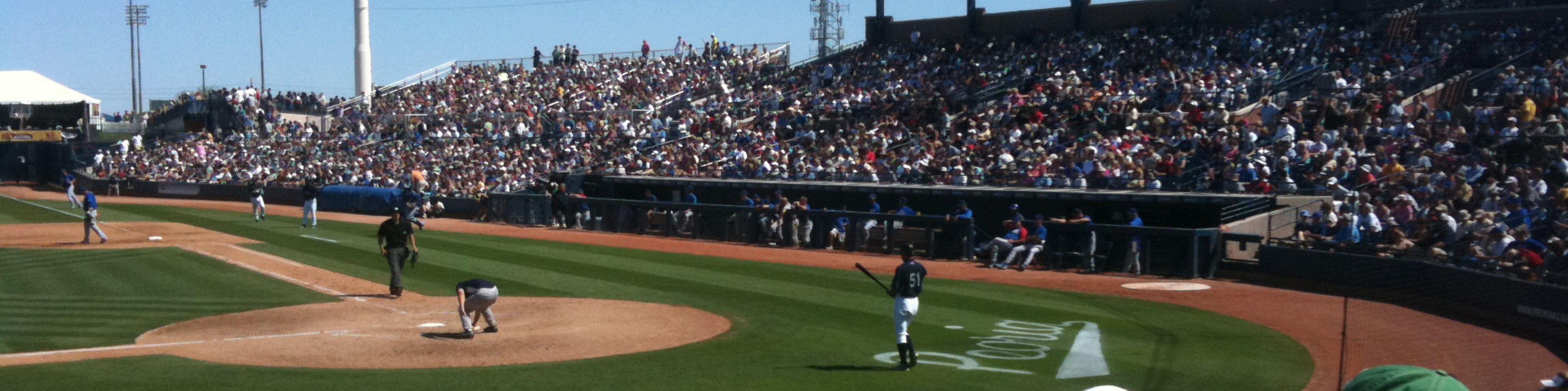}
  \caption{Seattle Mariners at Spring Training, 2010.}
  \Description{Enjoying the baseball game from the third-base
  seats. Ichiro Suzuki preparing to bat.}
  \label{fig:teaser}
\end{teaserfigure}

\received{30 July 2025}
\received[revised]{15 August 2025}
\received[accepted]{5 Sepetember 2025}

\maketitle

\section{Introduction}
In large-scale internet services, software changes — such as new feature rollouts, configuration updates, and patch fixes — are the driving force behind rapid business iteration. According to statistics, leading companies like Google execute over 10,000 changes per day. However, this high frequency of changes also brings significant risks: approximately 70\% of service failures are caused by faulty software changes, potentially resulting in tens of millions of dollars in economic losses. At Baidu, 54\% of service failures are attributed to changes. Such failures can lead to severe financial damage; for example, an erroneous software change led to a Facebook outage in 2021, which was estimated to have caused a loss of \$60 million. Therefore, an efficient and reliable change management process is crucial to ensure service stability.

A typical change lifecycle management process involves three core tasks: Erroneous Change Detection (ECD) — quickly identifying problematic changes; Failure Triage (FT) — categorizing failures and assigning them to the appropriate response teams; and Root Cause Change Analysis (RCCA) — precisely locating the specific cause that led to the failure.

Although both academia and industry have made progress in automating ECD (e.g., SCWarn\cite{scwarn}, Kontrast\cite{kontrast}), FT and RCCA still largely rely on engineers’ manual analysis. Existing practices face the following three key challenges:
\begin{enumerate}
    \item \textbf{Lack of Unification}: ECD, FT, and RCCA are often handled as isolated tasks by different tools or teams. This fragmentation leads to redundant work (such as repetitive feature extraction and data organization) and information silos, which greatly reduce the overall efficiency of failure diagnosis.
    \item \textbf{Lack of Interpretability}: In recent years, large language models (LLMs) have shown great potential for addressing these tasks in a unified manner due to their powerful natural language processing capabilities. However, the inherent “black-box” nature of LLMs makes their decision-making processes difficult to understand and trust. In high-risk scenarios like software change management, an AI system whose “reasoning” process cannot be explained is unlikely to be accepted.
    \item \textbf{Lack of Continuous Learning}: Online environments are constantly evolving, with new features being deployed continuously, giving rise to new failure patterns. Existing models lack effective feedback mechanisms, making it difficult for them to learn and adapt from new cases that occur daily. As a result, their performance degrades over time.
\end{enumerate}

\begin{figure}
    \centering
    \includegraphics[width=12cm, height=4cm]{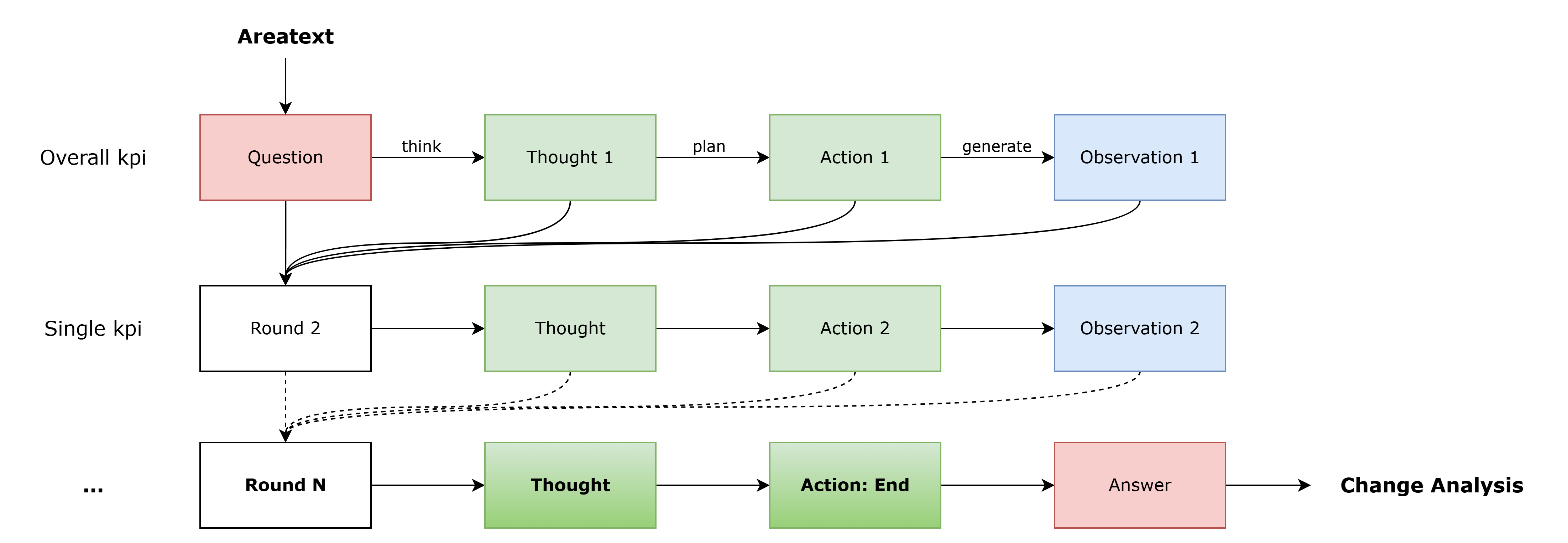}
    \caption{Chain of thought in Software Change analysis}
    \label{fig:cot-example}
\end{figure}

To this end, we propose SCELM to automate and streamline the critical stages of change management. SCELM integrates large language models (LLMs) with multi-task learning, consolidating these tasks into a single automated pipeline. Unlike existing methods that require separate manual workflows, SCELM simplifies the entire process — from detecting erroneous changes and classifying incidents to identifying root causes. SCELM employs a Retrieval-Augmented Generation (RAG) model to enhance the LLM’s ability to access and update operational knowledge, ensuring adaptability to evolving scenarios. Moreover, it uses reinforcement learning techniques to strengthen the model’s chain-of-thought reasoning figure ~\ref{fig:cot-example} and improve the interpretability of the generated structures. This unified approach increases efficiency, reduces the cost of manual inspection, and accelerates problem resolution, ultimately enabling more reliable and cost-effective service management.

To systematically address the challenges described above, we propose the SCELM (Software Change Evaluation and Lifecycle Management) framework and design its core reasoning method, SCoT (Software Change Chain-of-Thought):
\begin{enumerate}
    \item \textbf{Unified Automated Framework}: We present SCELM, the first framework to integrate ECD, FT, and RCCA into a unified LLM-based automated system, enhanced with RAG for dynamic knowledge retrieval.
    \item \textbf{Interpretable Reasoning Method}: We introduce SCoT, a Chain-of-Thought (CoT) and reinforcement learning (GRPO)-based method for change analysis, which significantly improves LLM reasoning ability and interpretability in specialized domains. We also design a new evaluation metric, CoTScore, to measure this.
    \item \textbf{Continuous Learning Mechanism}: We propose a lightweight feedback mechanism based on KTO, enabling the system to learn continuously from simple success/failure signals. This enhances the framework’s robustness in real-world deployments and improves the data transferability of LLMs in change scenarios.
    \item \textbf{Comprehensive Evaluation and Resources}: We validate our approach through extensive experiments on three real-world industrial datasets. Furthermore, we release the first CoT (Agent) dataset for change scenarios, providing the community with a valuable benchmark for future evaluation and research.
\end{enumerate}

\section{Preliminaries}
\subsection{Software Changes}
For large-scale online service systems such as social networks, online banking, and search engines, engineers need to perform frequent software changes to fix bugs, deploy new features, adapt to environmental changes, and improve software performance.
\begin{figure}
    \centering
    \includegraphics[width=10cm, height=4cm]{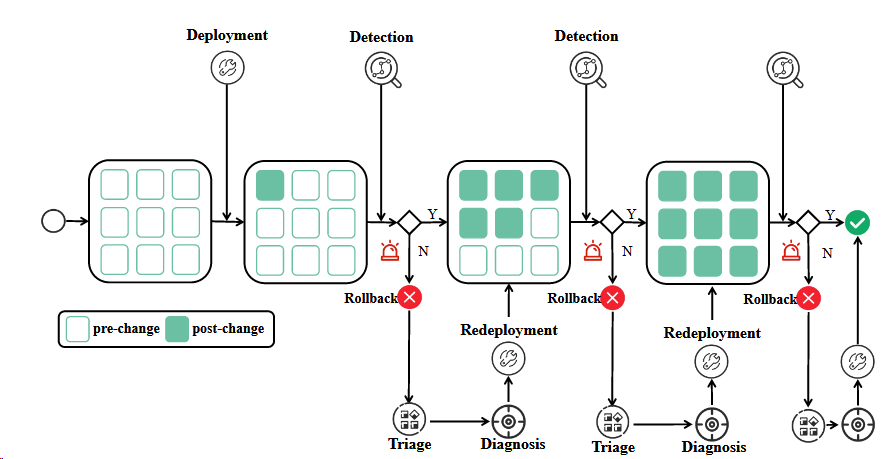}
    \caption{Lifecycle of software change}
    \label{fig:change-lifecycle}
\end{figure}
The software change lifecycle figure ~\ref{fig:change-lifecycle} consists of several stages: (1) Deployment: Developers deploy the new software version to a limited set of servers, virtual machines, or containers and monitor its status. (2) Detection: If an anomaly occurs, the deployment is halted and rolled back. (3) Classification: After an initial assessment, the incident is assigned to the appropriate engineering team. (4) Diagnosis: Engineers investigate the root cause, which often requires multiple rounds of communication. (5) Redeployment: Once the issue is resolved, the change is redeployed.

\subsection{Chain-of-thought in Large Language Model}
Large Language Models (LLMs), such as the GPT series, have demonstrated strong capabilities in text understanding and generation. In the AIOps domain, they are being applied to tasks such as log analysis, alert clustering, and incident report generation. By transforming operational data into natural language descriptions, LLMs have the potential to automate complex diagnostic tasks, greatly reducing the workload for engineers and improving problem resolution efficiency.

Chain-of-Thought (CoT) is a technique designed to enhance the complex reasoning capabilities of LLMs. Instead of directly outputting an answer, CoT generates a series of intermediate, logically coherent reasoning steps that simulate the human problem-solving process. For example, when performing mathematical calculations or reasoning tasks, CoT lays out each step of the computation or logical deduction rather than only providing the final result. This not only improves the accuracy of the final answer but also significantly enhances the interpretability of the model’s decision-making process. With CoT, we can better understand how an LLM, in the context of change scenarios, connects single-metric anomalies to inter-service calls and carries out downstream tasks such as anomaly detection and root cause analysis — which is crucial for AIOps scenarios that rely heavily on trustworthy model-driven decisions.

\subsection{Reinforcement Learning from Human Feedback}
Reinforcement Learning from Human Feedback (RLHF) is a key technique for fine-tuning the behavior of large language models (LLMs). The core idea is to enable the agent (i.e., the LLM) to interact with its environment and learn the optimal strategy by receiving feedback that reflects human preferences. In each interaction, the agent selects an action based on its current state (e.g., generating a piece of text); the environment then returns a signal (or feedback provided by humans) along with a new state. By continuously exploring and optimizing to maximize cumulative positive feedback, the agent gradually learns to make the best decisions in an uncertain environment.

When selecting an appropriate RLHF strategy, we primarily considered the following factors:
\begin{itemize}
    \item \textbf{Adaptability to long CoT tasks}: In diagnostic tasks, the LLM needs to generate detailed, multi-step reasoning processes.
    \item \textbf{Ease of data collection}: In practical operations scenarios, obtaining high-quality human feedback is often expensive and time-consuming.
    \item \textbf{Training efficiency and stability}: We expect the chosen algorithm to effectively improve model performance while ensuring computational efficiency and stability during training.
\end{itemize}

In our research, we employ two more advanced variants of RLHF:
\begin{enumerate}
    \item \textbf{GRPO (Group Relative Policy Optimization)} for enhancing the model’s CoT capability in change scenarios. GRPO is a policy optimization algorithm that does not require an explicit value estimation. Instead, it updates the policy by comparing the relative quality of a set of candidate outputs. This approach is well-suited for generative tasks, as it directly learns from human preferences over generated text.
    \item \textbf{KTO (Kahneman-Tversky Optimization)} for managing feedback on generated outputs. KTO is a lightweight alignment algorithm that does not require pairwise preference data — it only needs a judgment on whether a single output is “good” or “bad.” This simplified feedback mechanism makes it easier to deploy and scale in real-world applications, as collecting single-point good/bad labels is much more practical than gathering complex preference rankings.
\end{enumerate}

\section{Motivation}
To address the issues mentioned above, our research is motivated by two core insights: the need for unified data processing and the pursuit of interpretability in the reasoning process.
\subsection{Unified downstream works}
The primary motivation stems from the challenges posed by the fragmented state of current change management processes. The core of software change management consists of a series of interdependent downstream tasks: anomaly detection, fault classification, and root cause localization.

In current practice, these three tasks are often handled by different, isolated tools, resulting in workflow disruptions and information silos. Integrating them into a seamless, automated process fundamentally requires a unified “information carrier.” However, these tasks rely on diverse data sources and heterogeneous formats — including structured performance metrics, semi-structured service logs, and unstructured incident reports. Existing approaches that attempt to unify this data often do so at the cost of losing critical semantic information.

Therefore, our motivation is to create a unified framework built upon a novel data representation approach that transforms multi-source heterogeneous data into semantically rich “domain text” natively understandable by LLMs. This provides a solid foundation for bridging and unifying all downstream tasks.

\subsection{Reasoning analysis process}
Even if the data challenge is addressed, the “hallucination” and lack of interpretability of LLMs in critical decision-making tasks remain major barriers to their application in operations and maintenance. A simple conclusion is far from sufficient; operations engineers need to understand why the model arrives at a given conclusion. This drives our second core motivation: to build a transparent, trustworthy, and optimizable reasoning engine.

\begin{enumerate}
    \item \textbf{Chain-of-Thought (CoT) as the foundation of interpretability}: Inspired by CoT techniques, which simulate human reasoning by generating intermediate steps, we believe this is key to achieving interpretability in change analysis. A model capable of outputting its complete reasoning logic — “Because I observed an increase in metric A, and log B appeared, combined with change C, I conclude that the root cause is D” — will be far more credible and useful than a traditional black-box model.
    \item \textbf{RAG as dynamic knowledge augmentation}: In live environments, new issues continually emerge, and a model’s internal, parameterized knowledge alone is insufficient to handle all scenarios. Therefore, we incorporate Retrieval-Augmented Generation (RAG) to dynamically retrieve similar historical cases during reasoning, providing the model with up-to-date and relevant external knowledge — which is crucial for addressing rare failures.
    \item \textbf{Reinforcement learning as a reasoning optimizer}: Standard CoT does not guarantee that each generated reasoning step is optimal or entirely correct. We introduce reinforcement learning (e.g., GRPO and KTO) to guide and refine the model’s reasoning strategy through reward mechanisms. This ensures that the reasoning process is not only transparent but also optimizable, enabling the model to continuously improve its analytical accuracy and logical rigor based on domain feedback.
\end{enumerate}

\subsection{Problem Statement}
To address the high cost of data annotation, the poor cross-enterprise transferability, and the lack of unified standards in current processes, we introduce large language models (LLMs) to simulate how engineers process change-related textual information. Although LLMs possess excellent language understanding and generation capabilities, their “black-box” nature results in limited interpretability, posing risks when maintaining generated examples in real-world scenarios.

Considering data confidentiality and the high costs associated with using commercial LLMs, we adopt open-source, smaller-parameter models (e.g., 7B scale) to balance processing speed and computational efficiency. During inference, we integrate a Retrieval-Augmented Generation (RAG) mechanism to dynamically retrieve information from external knowledge sources, enhancing the model’s performance on key change tasks such as ECD, FT, and RCCA. Prior to reasoning, we use the GRPO algorithm to domain-adapt and strengthen the model’s Chain-of-Thought (CoT) capability; after reasoning, we apply the KTO algorithm to optimize the generated outputs with feedback and continuously update the experience knowledge base to improve the system’s continual learning ability.

On this basis, we propose a unified LLM framework that simulates the cognitive chain-of-thought process of an Operation Change Engineer (OCE). This framework supports integrated understanding of multi-modal data, automatic extraction of actionable insights, and timely delivery of evaluative outputs. It significantly improves the interpretability of LLM-generated content, reduces the need for human intervention, and enhances the accuracy and responsiveness of fault management, effectively supporting intelligent change management requirements in real-time systems.

\section{Approach}
\subsection{Overview}
\begin{figure}
    \centering
    \includegraphics[width=12cm, height=5cm]{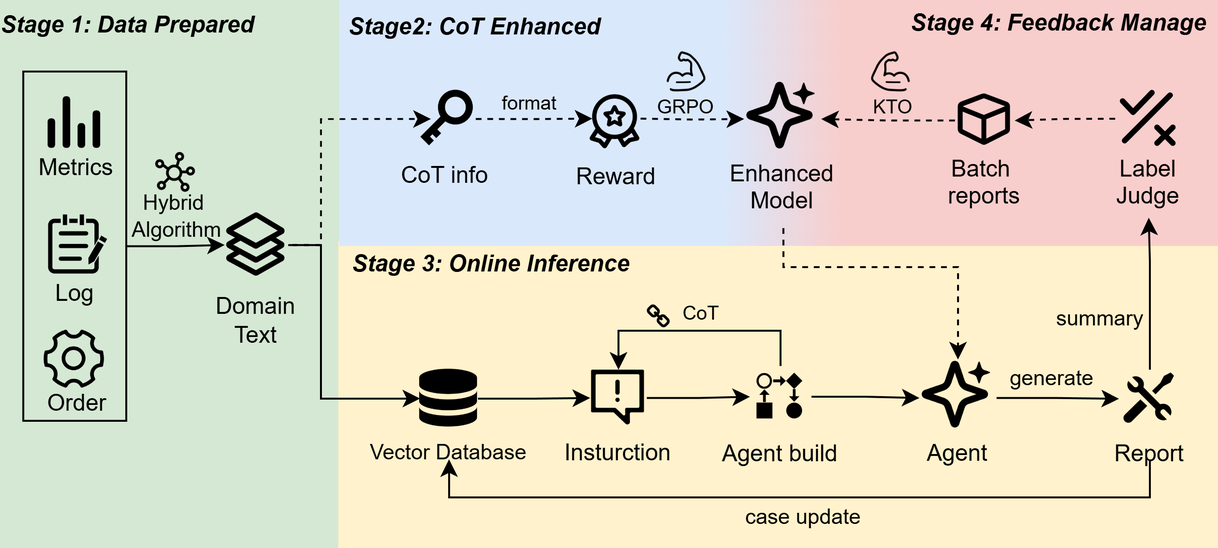}
    \caption{Overview of SCELM}
    \label{fig:overview-of-scelm}
\end{figure}
This study proposes a novel intelligent change analysis system aimed at improving the quality and interpretability of change ticket report generation. The system innovatively integrates offline reinforcement learning with online inference mechanisms, supplemented by multi-dimensional data unification and model enhancement strategies. The overall architecture is illustrated in the figure ~\ref{fig:overview-of-scelm}.

\subsubsection{Online and Offline Stage}
The system adopts a two-stage operational paradigm of offline reinforcement learning and online inference, aiming to balance improvements in model reasoning capabilities with efficient generation of real-time analysis reports.

\begin{itemize}
    \item \textbf{Offline Reinforcement Stage (Stage 2: CoT Enhanced)}:
 The core objective of this stage is to enhance the base model’s reasoning ability, enabling it to generate high-quality Chain-of-Thought (CoT) outputs. We innovatively introduce the Group Relative Policy Optimization (GRPO) reinforcement learning algorithm. By training on a preprocessed dataset, GRPO effectively optimizes the model’s decision-making policy, guiding it to form structured reasoning logic during inference. This significantly improves the interpretability and accuracy of the generated results. The model’s output includes not only the final inference but also detailed CoT information, providing rich intermediate steps for subsequent analysis.
    \item \textbf{Online Inference Stage (Stage 3: Online Inference)}:
 In this stage, the model enhanced through offline reinforcement is applied to actual change analysis tasks. The model first consults a Retrieval-Augmented Generation (RAG) vector knowledge base containing historical experience data. These historical cases serve as an external knowledge source, supplying rich contextual information and patterns of success/failure cases, which effectively mitigate the large model’s limitations in domain-specific knowledge. Driven by the powerful reasoning ability imparted by GRPO, the model performs in-depth analysis of input change events in linear time, ultimately generating detailed change ticket reports. The entire inference process is transparent and supported by CoT information, ensuring traceability and interpretability of the analysis results.
\end{itemize}

\subsubsection{Three core modules}
To realize the aforementioned two-stage operational paradigm, the system is further refined into three core modules, each responsible for key functionalities.

\begin{itemize}
    \item \textbf{Multi-Dimensional Data Unification Module (Stage 1: Data Prepared)}: This module focuses on unifying heterogeneous raw data into domain text modalities that the large model can process, thereby endowing the system with strong multi-modal processing capabilities. Specifically:
        \begin{itemize}
            \item For metric data, we employ pattern matching algorithms for structured extraction and transformation to capture trends in key performance indicators.
            \item For log data, we use the Drain algorithm for log template matching and parsing, converting unstructured logs into comprehensible event sequences.
            \item For order (ticket) data, we directly extract structured information.
        \end{itemize}
    All processed data are ultimately consolidated into a unified domain text representation, which serves as the input for downstream large models, ensuring completeness and consistency of information.
    \item \textbf{Model Enhancement Module}:This module is critical to improving the system’s intelligence and is divided into two main parts:
        \begin{itemize}
            \item \textbf{GRPO Enhancement}: As described earlier, through the GRPO reinforcement learning algorithm, we train the model on carefully constructed datasets to master complex reasoning abilities. This enables the model, when generating change analysis reports, not only to produce conclusions but also to present clear and logical reasoning processes, greatly enhancing result interpretability.
            \item \textbf{KTO Enhancement (Stage 4: Feedback Manage)}: To enable the model to learn from feedback, we introduce the Knowledge Transfer Optimization (KTO) algorithm. KTO facilitates continuous iteration and optimization of model knowledge by relearning from failed analyses and some poorly performing successful cases. The feedback management module (Label Judge) is responsible for manually or semi-manually evaluating generated reports and fine-tuning the model based on feedback signals (e.g., relevance feedback), thereby achieving continuous improvement and self-enhancement.
        \end{itemize}
    \item \textbf{Change Analysis Generation Module (Stage 3: Online Inference)}: This module constitutes the system’s final output stage. With the assistance of historical experience stored in the RAG vector knowledge base and the dual-enhanced inference capabilities from GRPO and KTO, the system efficiently and accurately generates change analysis reports. These reports not only provide deep insights and root cause analyses of change events but also include the overall chain-of-thought reasoning process (expressed via CoT), making the analysis more transparent and the conclusions more persuasive. The final reports guide subsequent decision-making and problem resolution, and through a “case update” mechanism, they feed back into the RAG vector knowledge base to form a closed knowledge loop.
\end{itemize}

\subsection{Unified Representation of Multimodal Data}
This module aims to enhance the identification and classification capabilities of fault points by analyzing data from Engineering Change Detection (ECD). It extracts anomaly-related information and change ticket data, providing targeted textual summaries for each case. The framework of this module is illustrated in the figure .

\begin{figure}
    \centering
    \includegraphics[width=12cm, height=5cm]{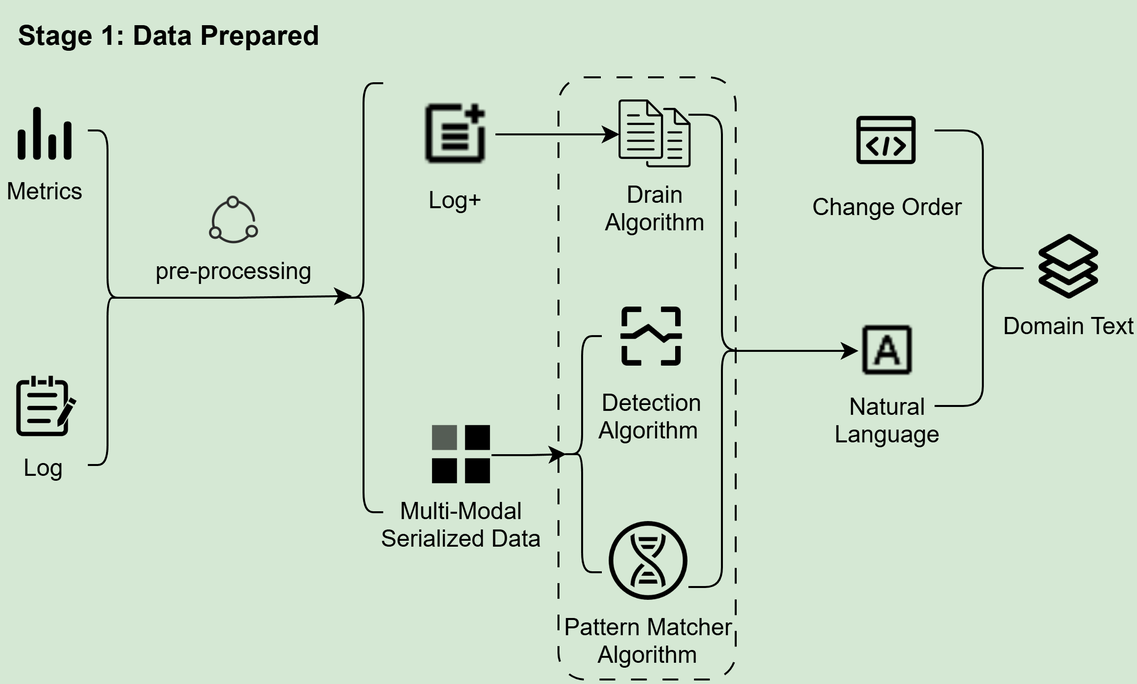}
    \caption{Multimodal unified process}
    \label{fig:unified_multimodel}
\end{figure}

\subsubsection{Multimodal data serialization and log processing}
To handle the various types of data involved in change evaluation, we serialize multimodal data (metrics and logs) into time series, as shown in figure ~\ref{fig:unified_multimodel} Specifically, metric data is preprocessed using standard normalization techniques, while log data is processed following the method in SCWarn. Concretely, we parse log templates using the Drain algorithm and calculate the frequency of the parsed templates, thereby converting logs into time series.

Typically, logs can be matched to existing templates. However, in erroneous change scenarios, the number of new logs increases significantly, resulting in new log templates that cannot be matched to existing ones. The SCWarn method addresses this by assessing the number of new logs unmatched to any template. Nevertheless, our observations in real-world scenarios indicate that new log templates caused by changes often contain important semantic information, which is crucial for fault diagnosis and root cause analysis.

As illustrated in the figure \ref{fig:change-order} below, L5 and L6 represent new log templates generated after the change. Although these logs cannot be matched to existing templates, they provide key fault insights — L5 indicates system failure, while L6 indicates business failure. Semantically, these logs point directly to the cause of the fault. Focusing solely on the quantity of new logs (as SCWarn does) would cause us to miss this vital semantic content. After preprocessing, these logs are converted into time series and integrated with metric data. Given the rich semantics of new log templates, they are represented in natural language form to enable collaborative processing by large language models (LLMs).

\begin{figure}
    \centering
    \includegraphics[width=12cm, height=4cm]{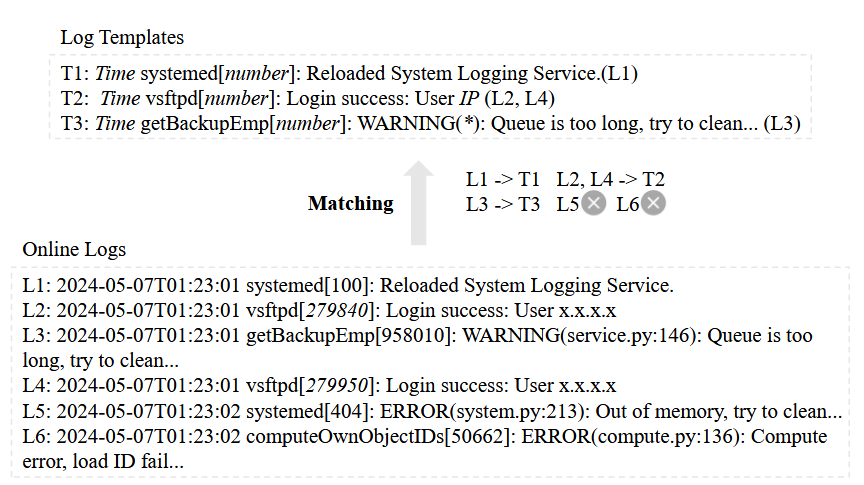}
    \caption{A typical example of change order}
    \label{fig:change-order}
\end{figure}

\subsubsection{Enhanced identification and failure triage}
\begin{figure}
    \centering
    \includegraphics[width=12cm, height=3cm]{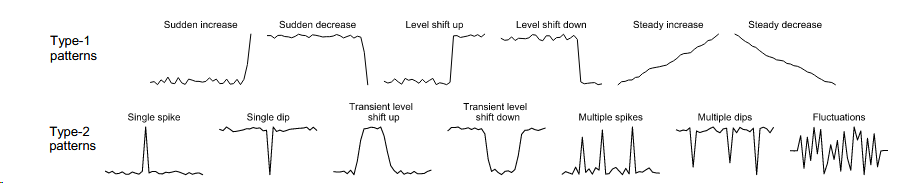}
    \caption{Patterns in time series}
    \label{fig:patterns}
\end{figure}

Anomalies in time series data can be classified based on their duration and shape. When anomalies are detected, operations engineers (OCEs) typically compare them with historical events to understand their nature and derive troubleshooting recommendations. We enhance this process by using a pattern matcher to define graphical rules and patterns for identifying the shapes of anomalies. This approach helps classify anomalies as either transient spikes or persistent issues.
For example, a “single spike” in response time may indicate a quickly resolved temporary problem, while a “steady increase” in memory usage could signal a more persistent issue. By classifying anomalies in this way, we can filter out irrelevant noise and focus on the most critical problems. As shown in figure ~\ref{fig:patterns}, these anomaly shapes are converted into natural language descriptions via the pattern matcher, making it easier for large language models to process and interpret the results.

\subsubsection{Integration of abnormal detection and change order}
We integrate multimodal serialized data with change ticket information to generate a domain text that consolidates the data from the previous steps and encapsulates the key insights needed by LLMs to produce accurate change evaluations.

The purpose of the domain text is to structure diverse data types—such as time series metrics, log data, and change ticket information—in a manner understandable by LLMs, thereby facilitating more precise change analysis. The domain text includes the following elements:
\begin{enumerate}
    \item Change Ticket Records: Including change ticket ID, affected services, change type (e.g., configuration change), and start and end times of submission and analysis.
    \item Identified Anomaly Timestamps: Indicating when anomalies were detected in the system, linking changes to observed issues.
    \item Anomaly Classification and Metric Descriptions: Defining anomaly types and describing related metrics.
    \item Pre- and Post-Change Metric Comparison: Comparing key metrics before and after changes and summarizing their impacts.
    \item Detailed Metric Comparison and Findings: Providing comparisons of individual metrics (e.g., max, min, average) before and after changes, along with result summaries.
    \item Descriptions of New Log Templates: Describing new log templates created after changes that may highlight new faults or variations.
\end{enumerate}

Historical change tickets typically contain complete information, including the status of past changes, whether errors occurred, fault types, root causes, and corresponding resolutions. These historical data are invaluable for constructing a historical experience knowledge base that informs future change analyses. Conversely, online change tickets provide real-time change-related information, helping to streamline the acquisition of multimodal data and making real-time evaluation generation more convenient.

By integrating these data points into coherent domain text, we ensure that LLMs receive structured and comprehensive inputs, enabling them to generate accurate results for tasks such as Erroneous Change Detection (ECD), Failure Triage (FT), and Root Cause Change Analysis (RCCA).

\subsection{Enhanced CoT with Software Knowledge}
\begin{figure}
    \centering
    \includegraphics[width=12cm, height=4cm]{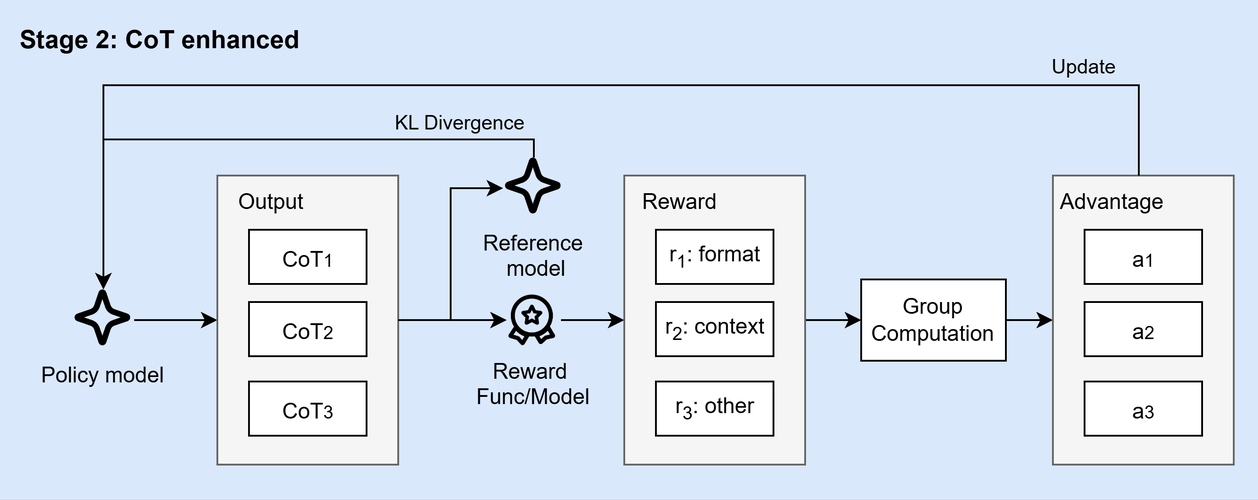}
    \caption{RLHF with domain knowlege}
    \label{fig:flhf}
\end{figure}
\subsubsection{Model the change management as a reinforcement learning task}
In the SCoT framework, anomaly detection, fault classification, and root cause localization together form an automated diagnostic chain focused on issues caused by changes. Although multimodal data (such as metrics and logs) have been serialized into unified domain text in the preceding modules, the reasoning chain still requires large language models (LLMs) to possess precise understanding and error correction capabilities for dynamically changing and structurally complex semantic content. To this end, we model fault analysis tasks in the change scenario as a Reinforcement Learning (RL) problem, enabling the model to continuously optimize its reasoning path through feedback-driven learning.

In this setup, the LLM is regarded as a policy agent, whose input is the uniformly preprocessed domain text (including metric trends, log patterns, new and old change contexts, etc.), and whose output consists of structured anomaly annotations, fault type identifications, and root cause localization suggestions. By designing a reward function that incorporates dimensions such as accuracy, confidence, and consistency, RL helps the model adjust its reasoning strategy in complex cases—particularly strengthening its ability to “self-correct from anomalies”—thereby improving the stability of reasoning and the quality of responses.

\subsubsection{Exploration of the Thought Chain Based on GRPO}

To enable SCoT to more effectively identify issues in change-induced reasoning tasks, we introduce the reinforcement learning optimization strategy GRPO.

\textbf{Group Relative Policy Optimization (GRPO)}: GRPO is an efficient policy optimization algorithm focused on enhancing the model’s fault detection and reasoning capabilities. By using a reward function to provide differentiated feedback to reasoning paths, GRPO encourages the model to explore more interpretable “chain-of-thought” structures, thereby uncovering latent correlations between anomaly patterns and potential faults. Within SCoT, the model leverages GRPO to compare multiple candidate reasoning paths, selecting more insightful causal chains and generating more targeted reports. This method does not rely on a critic network and is well-suited for real-world operations scenarios characterized by high feedback noise and inference ambiguity [3].

Therefore, GRPO enhances the model’s ability to actively construct chain-of-thought reasoning and identify critical fault points under multi-path conditions, significantly improving SCoT’s decision-making performance when handling complex anomalies in change scenarios.

\subsubsection{Enhanced interpretability and transparency in abnormal detection}
The reinforcement learning mechanism in SCoT not only improves the model’s prediction accuracy along the fault chain but also significantly enhances the structural transparency of its reasoning process. The stepwise reward strategies of GRPO and KTO encourage the LLM to form clear causal reasoning chains: from anomaly detection triggered by metric deviations, to fault classification supported by matched log templates, and finally to root cause localization combined with change records.

The model-generated reasoning paths visually express the basis for each decision point, facilitating review and decision-making by OCE engineers. Especially in sensitive operations such as canary releases, KTO’s conservative policy optimization guides the model to generate primary and backup suggested paths, ensuring the outputs are sufficiently interpretable and practically actionable. Overall, GRPO and KTO collaboratively enhance SCoT’s chain-of-thought expression ability, failure feedback utilization efficiency, and interpretability, making it more aligned with real-world engineering decisions when handling complex change-related faults.

\begin{figure}
    \centering
    \includegraphics[width=12cm, height=5cm]{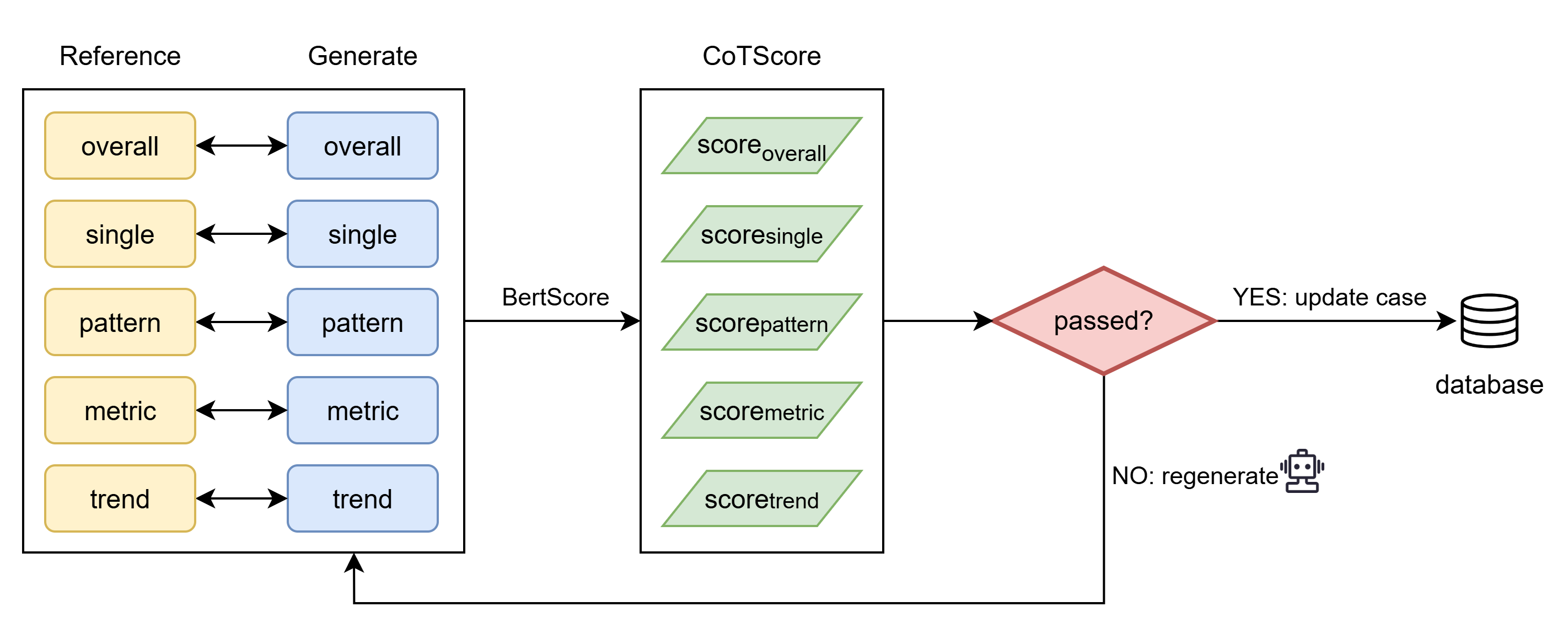}
    \caption{Evaluation process in CoTScore}
    \label{fig:CoTScore}
\end{figure}

Inspired by the BertScore algorithm, we designed an evaluation metric called CoTScore, as shown in figure ~\ref{fig:CoTScore}, based on the thinking patterns of change analysis. It divides the model’s generated analysis of change content into different modules and scores them against corresponding historical experience references, taking into account the weights of each module. An overall CoTScore is calculated; if it passes a set threshold, the result is added to the database. Otherwise, a rewriting agent is used to regenerate the output.

\subsection{Online inference augmented with historical experience}
\begin{figure}
    \centering
    \includegraphics[width=11cm, height=5cm]{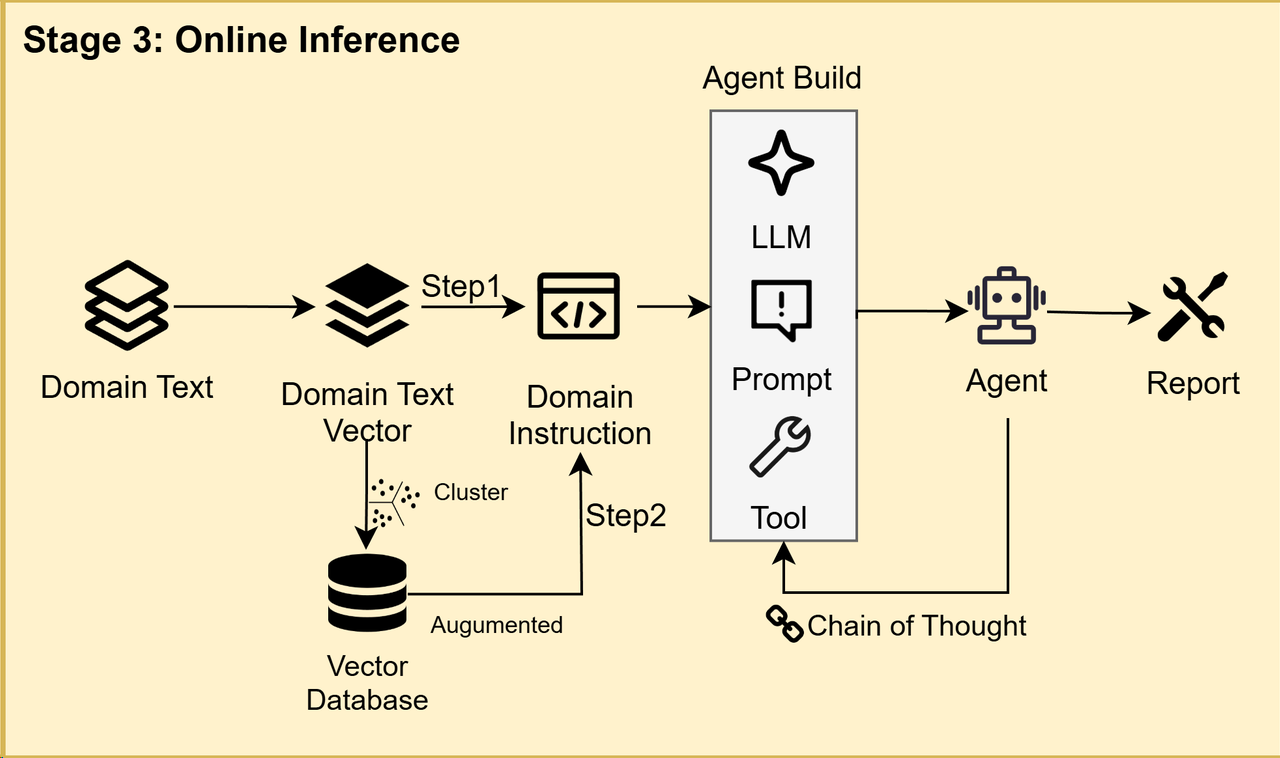}
    \caption{Online inference with RAG}
    \label{fig:online-inference}
\end{figure}

To ensure the model has real-time responsiveness to change events during actual operation, SCoT introduces an online inference mechanism , shown in ~\ref{fig:online-inference} , that integrates a historical experience knowledge base, combining Retrieval-Augmented Generation (RAG) technology with structured reinforcement reasoning capabilities to optimize responses to new anomalies.

\subsubsection{Hitstorical experience embedded enhancement}
When the system detects new anomalous metrics or log patterns, it first retrieves the most relevant historical change cases from the offline constructed RAG knowledge base. This knowledge base encompasses multidimensional structured corpora, including fault labels, anomaly patterns, root cause paths, and corresponding mitigation measures.

\subsubsection{Reasoning mechanism process}
\begin{enumerate}
    \item Problem Expression Construction: Convert the current change context into a structured query.
    \item RAG Retrieval: Match and extract analysis reports of historically similar issues.
    \item Inference Input Construction: Concatenate the retrieved case summaries with the current problem as the input prompt.
    \item Structured Reasoning Generation: Use the reasoning model trained with GRPO to generate a complete analysis chain, including predicted labels, explanation paths, and mitigation suggestions.
\end{enumerate}

\begin{figure}
    \centering
    \includegraphics[width=12cm, height=5cm]{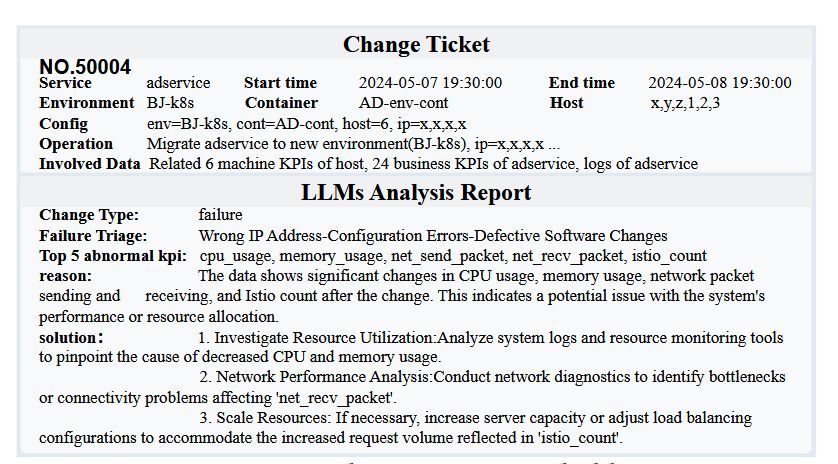}
    \caption{Change Analysis by SCELM}
    \label{fig:scelm-analysis}
\end{figure}

This approach combines structured knowledge with deep reasoning, achieving dynamic response and high-precision understanding of change events. As shown in figure ~\ref{fig:scelm-analysis}.

\subsection{Case feedback management}
\subsubsection{Database update}
To ensure that the RAG retrieval database continuously reflects the latest operational knowledge and fault cases, the system incorporates a Case Feedback Management module to dynamically update and maintain the database. This module collects and analyzes feedback from change analysis reports generated during the online inference phase, including success and failure annotations, to automatically identify knowledge points that need supplementation or correction. Based on this feedback, the system regularly integrates verified new fault cases, anomaly patterns, root cause analysis paths, and mitigation strategies into the RAG database, achieving continuous evolution of the knowledge base.

 This closed-loop update mechanism not only improves the relevance and accuracy of retrieval but also enhances the model’s responsiveness to novel or complex faults, driving lifelong learning and optimization of the entire intelligent change management system.

 \subsubsection{Feedback-driven Optimization Based on KTO}
To enable SCoT to receive feedback and continuously self-improve, we introduce the reinforcement learning strategy KTO:

\textbf{Kahneman-Tversky Optimization (KTO)}: KTO focuses on the model’s absorption and utilization of failure experiences. By assigning significant weight to feedback from failed cases, KTO enables the model to identify and avoid reasoning paths that have been historically proven to be high-risk or misjudged in change scenarios. This method guides the model to form a self-correction mechanism, allowing iterative refinement of reasoning strategies during ongoing operation. KTO relies solely on binary success/failure feedback, making it efficiently adaptable to real rollback records and diagnostic labels found in change logs [4]. KTO enhances the model’s sensitivity to failure feedback and its capacity for self-improvement. When combined with GRPO for joint optimization, it significantly boosts SCoT’s robustness across diverse problem environments.

\section{Evaluation}
\subsection{Experimental Setup}
\begin{itemize}
    \item Datasets: yunzhanghu, kontrast, bytedance
    \item Baseline Approach
    \item Evaluation Metric
    \item Deployment: environment
\end{itemize}

\subsection{RQ1: Performance in ECD, FT, RCCA}
\begin{figure}
    \centering
    \includegraphics[width=12cm, height=5cm]{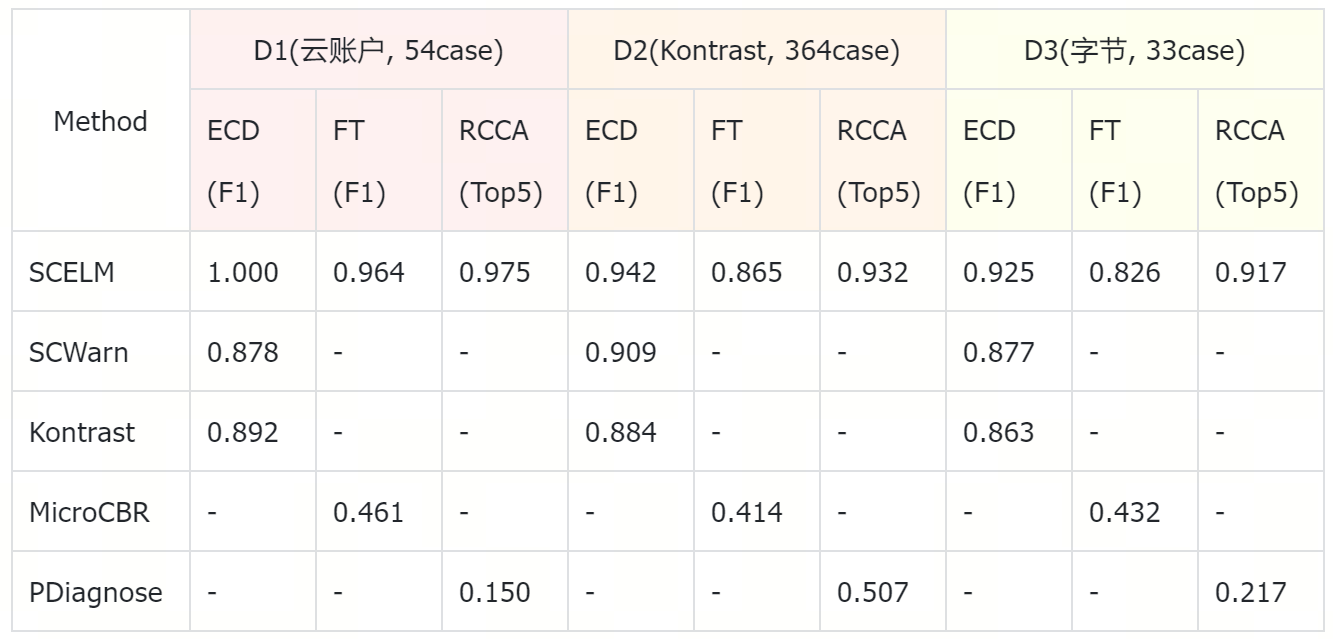}
    \caption{Overall result}
\end{figure}
\begin{itemize}
    \item For Exceptional Change Detection (ECD): SCELM demonstrated outstanding detection capability, achieving F1-scores of 1.000, 0.942, and 0.925 on datasets D1, D2, and D3, respectively. By comparison, baseline methods SCWarn, Kontrast, Lumos, Funnel, and Gandalf had average F1-scores of 0.8935, 0.888, 0.868, 0.869, and 0.8685, respectively; these methods typically focus on single-modal data or straightforward temporal comparisons. The superiority of SCELM lies in its ability to synthesize multimodal signals through a unified domain text representation—for instance, simultaneously capturing spikes in CPU metrics and newly emerging “connection timeout” errors in logs. This comprehensive approach facilitates more accurate decision-making and effectively reduces false negatives and false positives.
    \item For Fault Classification (FT) and Root Cause Change Analysis (RCCA): SCELM achieved F1-scores of 0.964 and 0.865, along with Top1-scores of 0.775 and 0.879 on two datasets. In contrast, MicroCBR only achieved F1-scores of 0.461 and 0.414, while PDiagnose's AVG@5 scores were 0.150 and 0.507, which were notably lower than SCELM's Top1 performance. This performance gap likely arises because MicroCBR and PDiagnose have limited sensitivity to fine-grained changes during transition processes, failing to capture correlations between metrics and logs. Additionally, in PDiagnose, the analytical outcomes from each modality are sequentially influenced by the preceding modality, potentially causing cascading errors.
\end{itemize}

\begin{figure}
    \centering
    \includegraphics[width=12cm, height=5cm]{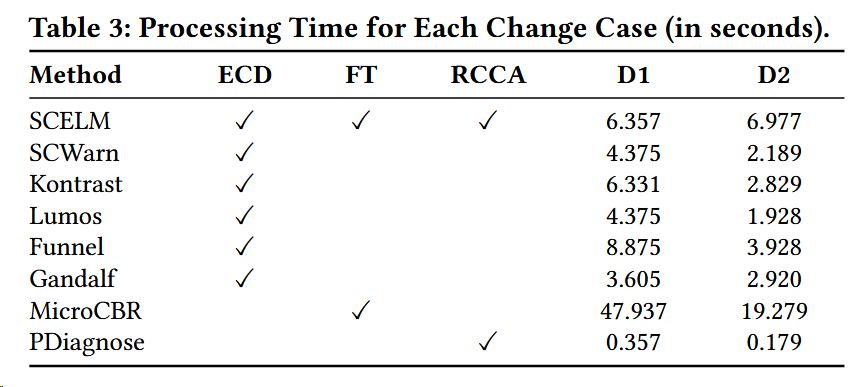}
    \caption{Running time}
\end{figure}

The table above compares the efficiency of SCELM with that of the baseline methods. SCELM outperforms the baselines by completing more tasks per change case within the same time window. Its approach mirrors the workflow of real-world operations engineers and ensures high performance through a lightweight design. Although it is slightly slower overall—for instance, SCWarn takes 4.375 seconds on D1, whereas SCELM takes 6.357 seconds—SCELM completes three tasks (ECD, FT, and RCCA) within that time, while SCWarn handles only ECD. This demonstrates SCELM's efficiency and its suitability for end-to-end change workflows. Overall, the results highlight the practical applicability of SCELM, showcasing its ability to efficiently perform real-time ECD, FT, and RCCA within a unified framework.

\subsection{RQ2: Performance of CoT}
\begin{figure}
    \centering
    \includegraphics[width=8cm, height=2cm]{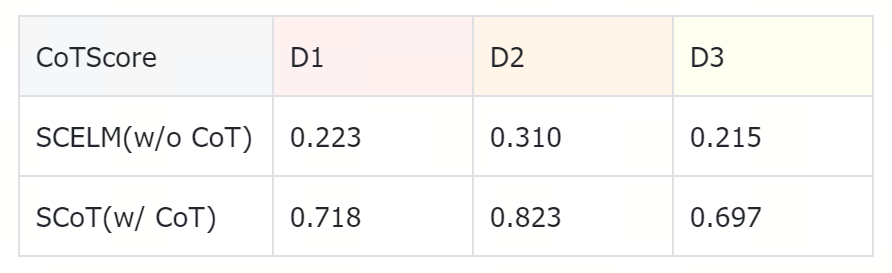}
    \caption{Performance with CoT}
\end{figure}
This experiment aims to evaluate the quality of the model’s reasoning process, rather than solely focusing on the accuracy of the final answers. To this end, we propose the CoTScore metric, which measures the alignment between the model-generated chain of thought and the expert diagnostic reasoning (serving as the reference standard).

The evaluation of CoT performance focuses on the overall reasoning trajectory, which progressively narrows from identifying system-wide anomalies to fine-grained, single-metric abnormalities. Experimental results show that without CoT, the overall CoTScore remains around 0.2, indicating that the analysis of changes tends to stay at a coarse-grained level. In contrast, the introduction of CoT leads to a significant 2–3× improvement in CoTScore. This demonstrates that SCoT not only reaches correct conclusions, but more importantly, generates a logically sound, clear, and expert-aligned reasoning process.

For example, without CoT, a vague response such as “It’s a CPU issue overall” might be produced (low CoTScore). In contrast, with CoT, the model can produce a structured explanation: Observation: CPU utilization spikes after the change. Analysis: The logs contain numerous “out-of-memory” errors, and the change involves a library known to have memory leak issues. Conclusion: The root cause is the memory leak introduced by the new library, leading to frequent garbage collection and consequently high CPU load (high CoTScore).

\subsubsection{The effect of SCoT in solving complex reasoning problems}
Theoretically, the introduction of reinforcement learning (RL) endows the SCoT model with the ability to autonomously explore chains of thought, thereby enabling it to tackle complex reasoning tasks more effectively. Specifically, traditional models typically rely on static reasoning paths or pre-defined rules, which constrains their performance when faced with novel and dynamic anomaly scenarios. In contrast, the SCoT model leverages the GRPO reinforcement learning approach, allowing it to adaptively select and adjust reasoning paths based on feedback during inference, dynamically exploring different decision sequences. This capability is particularly well-suited for complex tasks such as Root Cause Change Analysis (RCCA), which often requires multimodal data fusion, anomaly pattern recognition, and cross-modal reasoning.

From a cognitive perspective, the advantages of the SCoT model are twofold: first, it simulates the dynamic decision-making and exploratory behavior of real-world engineers when confronted with complex problems—a process that is inherently iterative and interactive; second, the model actively tests alternative hypotheses and leverages environmental feedback to progressively converge toward optimal solutions, thereby reducing the risk of misjudgment caused by the rigid reasoning pathways typical of conventional models.

Experimental evaluations further support the theoretical claims. On the RCCA task, the SCoT model significantly outperforms both its predecessor and other existing models in terms of F1-score, indicating not only more accurate root cause localization but also a substantial reduction in false alarms and missed detections. This performance improvement can be attributed to the reinforcement learning framework, which enables more thorough exploration of the reasoning space during training and allows timely strategy adjustment based on feedback, thus achieving higher reasoning accuracy and stability.

In conclusion, both theoretical analysis and empirical results confirm the substantial superiority of the SCoT model in handling complex reasoning tasks, with particularly strong performance in RCCA. This offers an efficient and reliable solution for automated anomaly management in complex systems and carries significant theoretical and practical implications.

\subsubsection{Correspondence between the CoT and plain text}
Through the quantitative evaluation provided by CoTScore, the SCoT model’s advantage in reasoning ability is prominently reflected at a macro level. Specifically, CoTScore measures the semantic alignment between each semantic segment of the model's output and the annotated reference with fine-grained precision. Evaluation results indicate that the overall semantic alignment of the SCoT model significantly surpasses that of traditional models that do not incorporate reinforcement-based chain-of-thought exploration, marking a qualitative leap in performance.

This substantial macro-level disparity highlights the SCoT model’s capacity to more comprehensively and deeply capture and interpret the underlying semantic structures of complex problems during the reasoning process, thereby enhancing the transparency and interpretability of its outputs. The fundamental improvement in semantic understanding and explanatory capability lays a solid foundation for the model’s reliable deployment in real-world complex reasoning tasks, and clearly underscores the critical role of reinforcement learning in elevating the reasoning competence of such models.

\subsection{RQ3: Lifelong learning Management}
\subsubsection{Cold start ratio of historical experience}

\begin{figure}
    \centering
    \includegraphics[width=12cm, height=6cm]{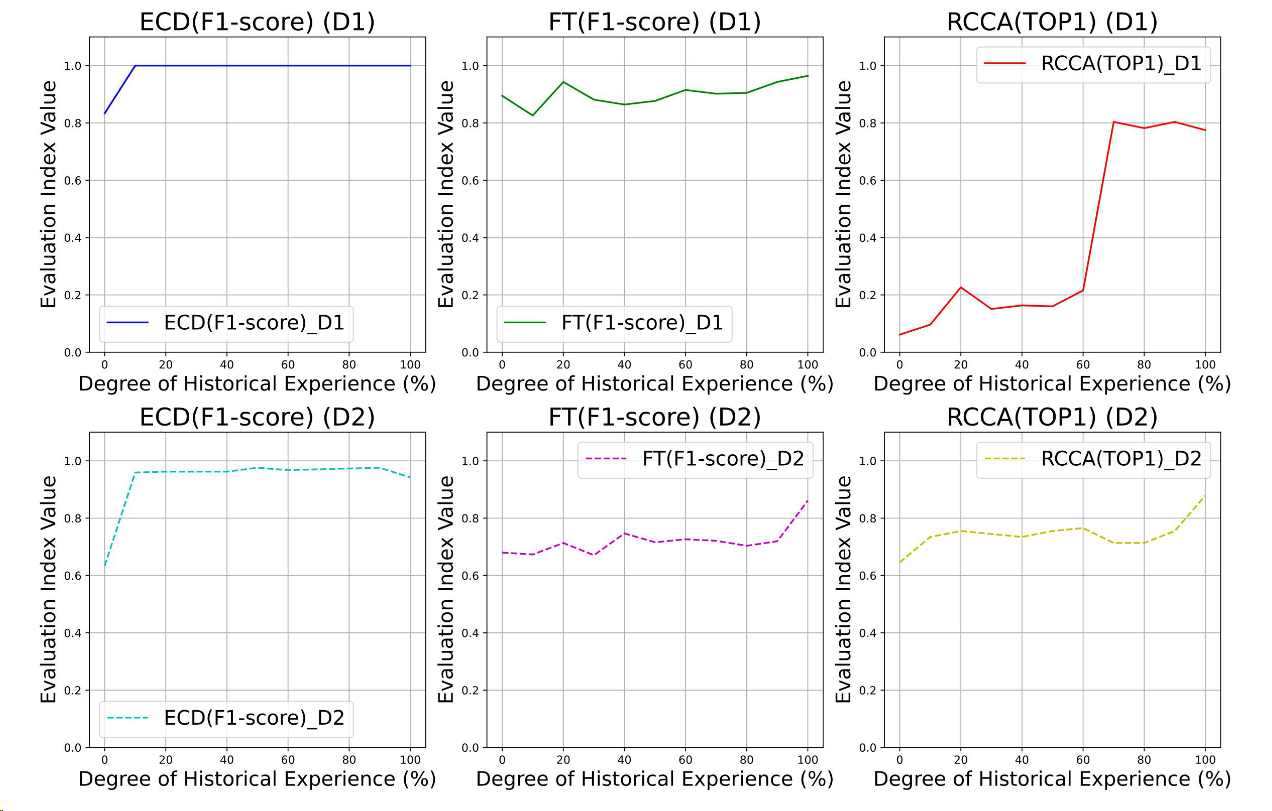}
    \caption{Code start ratio}
\end{figure}
We simulate the system’s learning process by varying the proportion of historical knowledge in the knowledge base, ranging from 0\% to 100\%.

We use the degree of domain expertise in the RAG component as a hyperparameter in SCELM to investigate how historical experience influences algorithm performance. This provides practical guidance for real-world deployment. Historical cases are randomly sampled at varying proportions from datasets D1 and D2. The experimental results are illustrated in Figure 10, where the solid line represents D1 and the dashed line represents D2.

\textbf{Cold start (0\% historical experience)}. Even without any historical knowledge (cold start), SCELM demonstrates reasonable performance, particularly in the ECD task.

\textbf{Increasing historical experience (up to 10\%)}. As the proportion of historical experience increases, performance improves steadily. However, beyond 10\%, further gains become marginal. This is likely because ECD focuses on anomaly detection, which heavily depends on data fluctuations and the physical characteristics of the metrics themselves.

\textbf{Impact on FT (beyond 10\%)}. For FT, improvements plateau around 20\% historical experience, with some slight declines observed in D1. This is attributed to the small number of cases in D1—sampling 10\% results in only two cases, which may either belong to the same category or vary significantly. Therefore, in practical deployment, it is essential to cover a broader range of cases even when dealing with a limited number of instances, to facilitate more effective algorithm learning.

\textbf{Impact on RCCA (beyond 70\%)}. For RCCA, improvements become negligible after reaching 70\% historical experience, particularly in D1. The larger case volume in D2 helps stabilize results, whereas the limited case diversity in D1 constrains the model’s learning until approximately 70\% coverage is reached. In real-world deployment, when historical experience is scarce, the algorithm can prioritize ECD and FT. Once the number of distinct cases exceeds around 20, RCCA performance improves significantly.

\subsubsection{Feedback error sample feedback ratio}

\begin{figure}
    \centering
    \includegraphics[width=12cm, height=3cm]{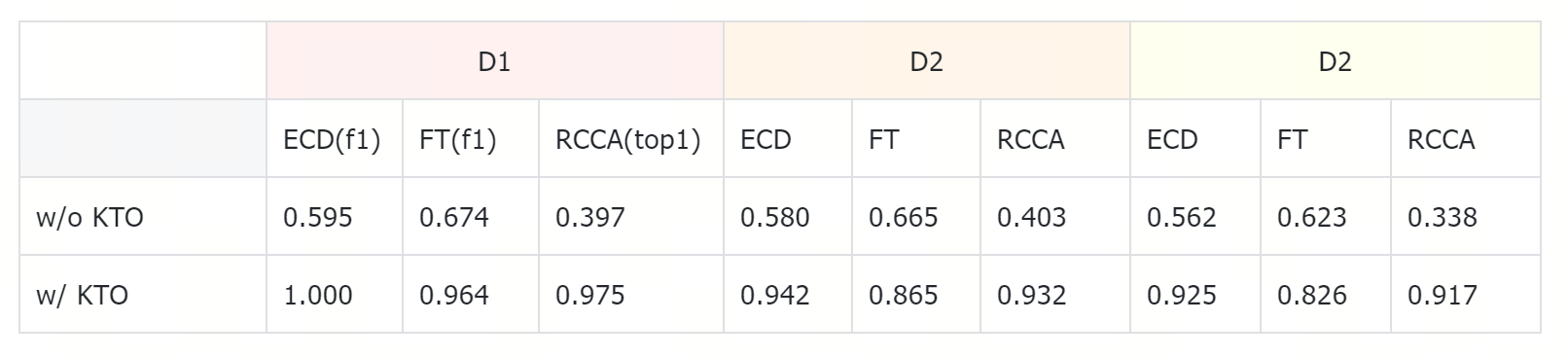}
    \caption{Feedback ratio of KTO}
\end{figure}
We simulate feedback learning from misdiagnosed cases through the KTO mechanism. Experimental results show that providing simple "good/bad" feedback on just a small number of failure cases can lead to a significant improvement in diagnostic accuracy when the system encounters similar issues in the future. This validates that the SCELM framework possesses lifelong learning capabilities, enabling it to continuously evolve and self-improve in production environments through minimal interaction with operations personnel.

\subsection{RQ4: Ablation Study}
\begin{figure}
    \centering
    \includegraphics[width=12cm, height=3cm]{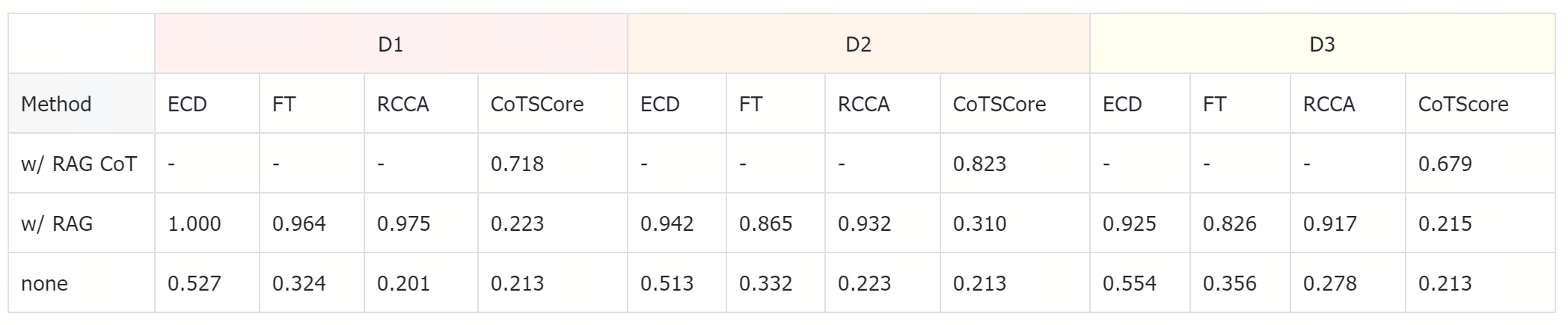}
    \caption{Ablation Study}
\end{figure}
\subsubsection{Ablation of RAG and CoT}
The "none" setting refers to directly using the LLM for generation without any external augmentation. Under this setting, ECD accuracy hovers around 50\%, indicating a state close to random guessing. The overall performance on FT anomaly classification and root cause localization is also poor, primarily due to the model's inability to grasp contextual information.

\textbf{Foundational Role of RAG}: When the RAG module is removed (as in the "none" version), the performance across all tasks deteriorates drastically, approaching that of random predictions. This strongly indicates that without the ability to dynamically retrieve domain knowledge from historical cases, the LLM lacks the contextual grounding necessary to solve professional operations and maintenance (O\&M) problems. Thus, RAG serves as the cornerstone of the entire framework’s effectiveness.

\textbf{Core Value of SCoT}: Comparing the w/ RAG and w/ RAG + SCoT variants reveals a crucial insight: without CoT, the model may still achieve high accuracy on downstream tasks, yet its CoTScore remains very low. This suggests that even when a standard RAG-augmented LLM can “guess” the correct answer, its reasoning process is disorganized and untrustworthy. The introduction of SCoT significantly boosts CoTScore—by several-fold—without sacrificing task accuracy. This demonstrates that SCoT’s core value lies in transforming the model’s “black-box” reasoning into a more transparent and interpretable “white-box” process, thereby enhancing both the reliability and explainability of the system while maintaining high performance.

\subsubsection{Ablation of methods}
\begin{figure}
    \centering
    \includegraphics[width=12cm, height=4cm]{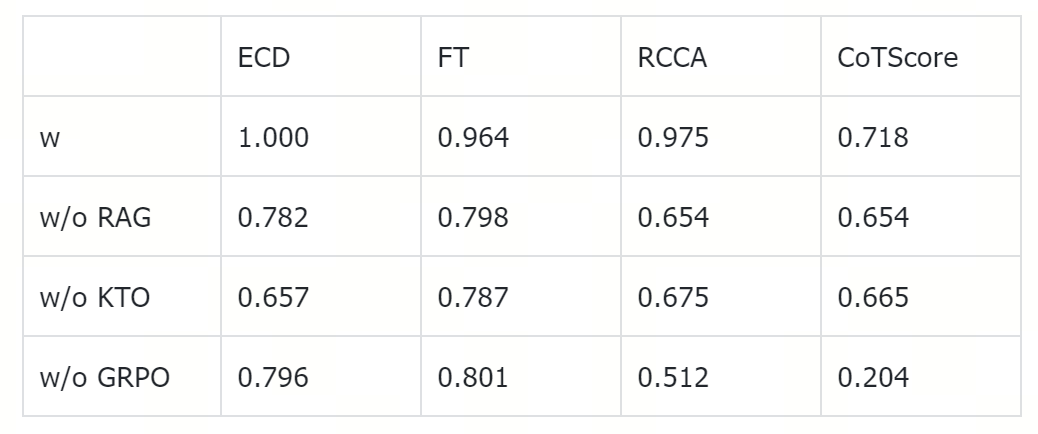}
    \caption{Ablation of Method}
\end{figure}
Through the quantitative evaluation enabled by CoTScore, the superiority of the SCoT model in reasoning capability is prominently demonstrated at a macro level. Specifically, CoTScore assesses the semantic alignment between each segment of the model’s output and the annotated reference with fine granularity. The evaluation results show that the overall semantic coherence of the SCoT model significantly surpasses that of traditional models without reinforced chain-of-thought exploration, marking a qualitative leap. This substantial macro-level gap reflects SCoT's enhanced ability to comprehensively and deeply capture and interpret the intrinsic semantic structures of complex problems, thereby improving the transparency and interpretability of its outputs. This fundamental advancement in semantic understanding and explanation capability provides a solid foundation for the model’s reliable application in real-world complex reasoning tasks, and underscores the critical importance of reinforcement learning in elevating model reasoning performance.

\textbf{How RAG Mitigates Hallucinations and Improves Change Expertise}: 
The ablation analysis of the RAG module further underscores its theoretical value in mitigating model hallucinations and enhancing domain expertise. Specifically, in the absence of RAG, the model exhibits a significant decline in its ability to capture contextual semantic associations and integrate external knowledge, leading to notable performance degradation in tasks such as anomaly detection (ECD) and fault classification (FT). Theoretically, RAG improves the model’s understanding of complex contexts and its real-time adaptability by retrieving relevant experiences and semantic fragments from an external knowledge base. This dynamic retrieval process substantially reduces the risk of hallucinated reasoning and enhances the credibility of model outputs.

\textbf{How GRPO Enhances Interpretability}: The ablation results of the GRPO method further elucidate its deep impact on model interpretability through reinforced chain-of-thought exploration. From a theoretical standpoint, GRPO enables the model to autonomously explore diverse reasoning paths during inference, allowing it to dynamically adjust decision sequences and evaluate the effectiveness of different reasoning trajectories in real time. When GRPO is removed, the model loses the flexibility to explore and adapt to unfamiliar problem spaces, resulting in a significant performance drop in root cause change analysis (RCCA). This absence of reasoning-path exploration directly reduces the model’s semantic alignment and interpretability, as reflected by a marked decline in CoTScore.

\textbf{How KTO Improves Lifelong Learning}: The ablation study of the KTO method clearly demonstrates its pivotal role in supporting the model’s lifelong learning capability. Theoretically, lifelong learning is especially crucial in dynamic domains such as software change management, where models must continuously adapt to evolving environments based on newly observed data. Without the KTO-based reinforcement learning mechanism, the model’s ability to autonomously learn from erroneous or suboptimal cases is severely impaired, resulting in a noticeable reduction in ECD performance. This highlights how KTO facilitates continuous strategy refinement through feedback-driven reinforcement, thereby enabling the model to sustain high performance in changing environments. It underscores the essential and irreplaceable value of lifelong learning in practical, real-world applications.

\subsubsection{Ablation of Key process Components}
\begin{figure}
    \centering
    \includegraphics[width=14cm, height=6cm]{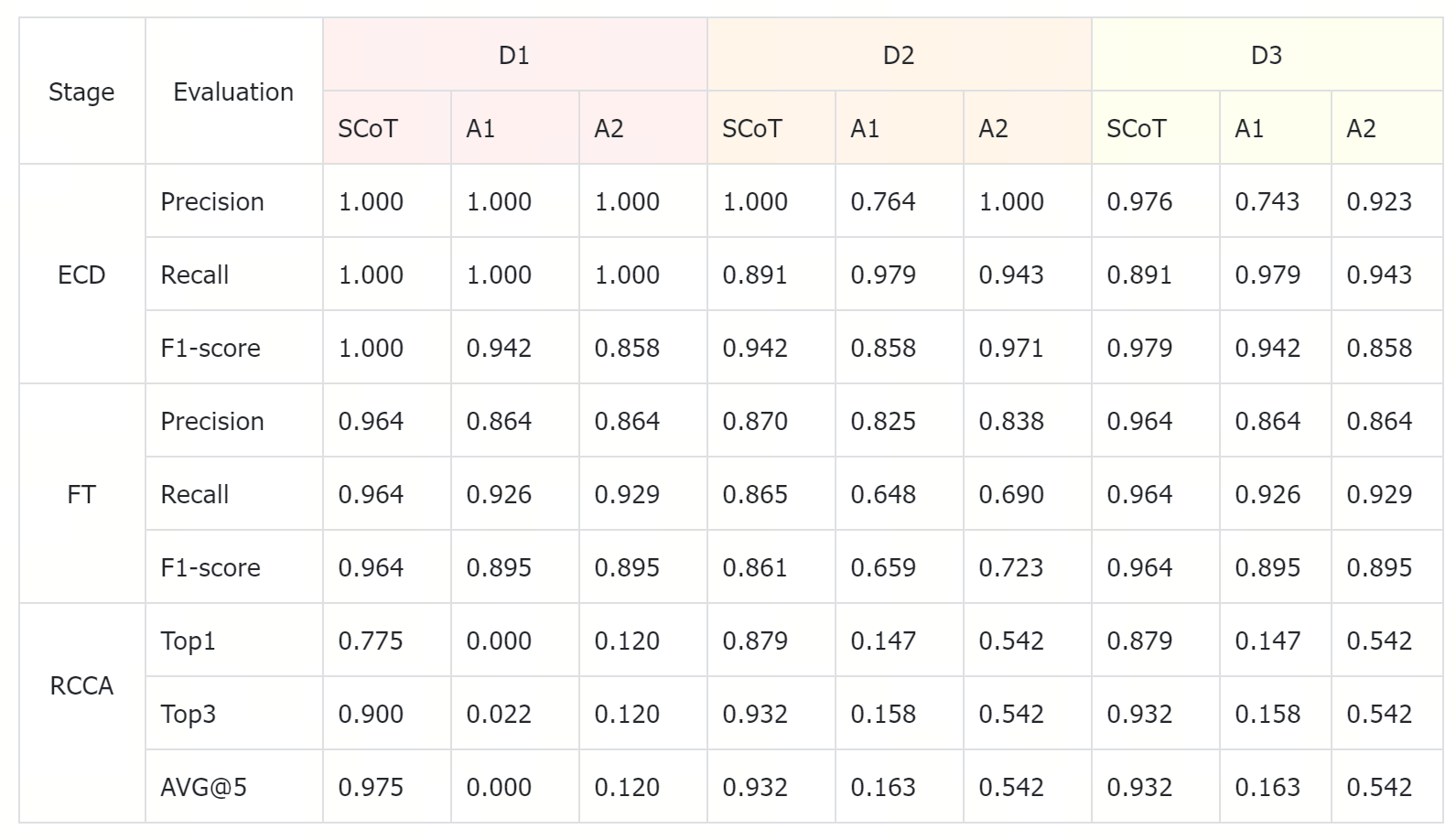}
    \caption{Ablation of modules}
\end{figure}
To validate the effectiveness of the two key components of SCELM—(1) the natural language description of data in the domain text, and (2) the detection algorithm for multimodal data—we conducted an ablation study. Two variants of SCELM were constructed: A1, which removes the data descriptions from the domain text, and A2, which omits the detection algorithm for multimodal data. The results for each variant are presented in Table.

\textbf{ECD Performance}. Experimental results show that removing the two key components—natural language descriptions and detection algorithms—has minimal impact on the ECD task. This can be explained by the fact that ECD is fundamentally an anomaly detection task, and LLMs are inherently capable of recognizing data fluctuations. In variant A2, even without using the detection algorithm, the model still performs satisfactorily using only the natural language descriptions. This suggests that the physical semantics of the data—as explicitly conveyed in natural language—carry significant value. In real-world scenarios, engineers often make judgments based on the physical meaning of data, with detection algorithms serving more as auxiliary tools.

\textbf{Impact on FT}. The impact on fault classification (FT) follows a similar trend to that of ECD. Nevertheless, SCELM still outperforms both variants overall. This is because SCELM performs classification primarily based on the physical semantics of the data—an approach consistent with how engineers classify faults in practice. Engineers typically rely on semantic understanding rooted in the physical characteristics of the data, and LLMs are well-suited to handle such semantic information. Therefore, even with certain components removed, SCELM maintains superior performance.

\textbf{Impact on RCCA}. In contrast, the RCCA task is much more sensitive to the removal of natural language descriptions. Without these descriptions, SCELM’s ability to identify root causes is significantly impaired. In the A1 variant, root cause detection is nearly nonexistent, with all Top-5 results returning null values—particularly in the D1 dataset. This indicates that without natural language context explaining the physical meaning of the data, SCELM struggles to infer root causes. When the detection algorithm is also removed, as in A2, the performance is not as poor as in A1, but still far from ideal. This suggests that the detection algorithm and natural language descriptions work synergistically in the RCCA task. In real-world settings, engineers typically combine detection algorithm outputs with their understanding of data semantics to determine root causes. Accordingly, the integration of both elements within SCELM enables the model to replicate this reasoning process and deliver more accurate root cause assessments.

\subsubsection{Different model}

\begin{figure}
    \centering
    \includegraphics[width=12cm, height=3cm]{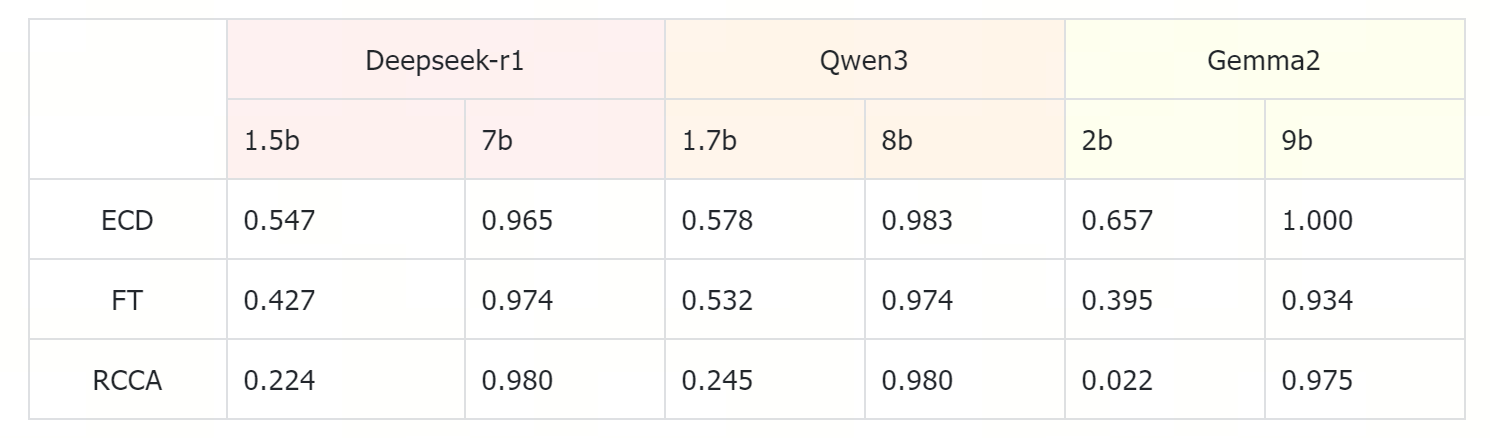}
    \caption{Hyperparameters of different models}
\end{figure}
This table summarizes the performance of three large language models—Deepseek-r1, Qwen3, and Gemma2—on three core tasks in change management: Exceptional Change Detection (ECD), Fault Classification (FT), and Root Cause Change Analysis (RCCA). The models span a range of sizes from 1.5B to 9B parameters.

Overall, performance consistently improves with increasing model size across all tasks, indicating that larger models possess stronger reasoning and accuracy capabilities in handling change management scenarios. Among the three, Gemma2 stands out as the top performer, especially in the ECD task, where its 9B version achieves a perfect F1-score of 1.000, demonstrating exceptional capability. Deepseek-r1 and Qwen3 exhibit comparable results, with both achieving high scores at medium-to-large scales (7B, 8B).

The RCCA task proves to be more challenging, with smaller models struggling to deliver satisfactory performance. However, larger models show substantial improvements—Gemma2-9B achieves an F1-score of 0.975, approaching the performance levels seen in ECD and FT.

These findings highlight the critical role of model scale and architecture in determining performance on change-related tasks. Notably, Gemma2 demonstrates strong potential for application in this domain. The results collectively underscore the value of large-scale language models in improving the accuracy and efficiency of automated software change management systems.

\section{Implementation}
The proposed SCELM system has been successfully deployed within the automated change management platform of a large-scale SaaS microservices infrastructure. As illustrated in Figure 11 (to be added; this can schematically depict SCELM’s integration point and workflow within the automation platform), when a new change is initiated, users submit the change request via the platform, which generates a change ticket and begins execution. SCELM continuously monitors the change process in real time, provides live status updates to users, and issues immediate alerts upon detecting erroneous changes. It can automatically trigger rollback operations and generate detailed change analysis reports to assist engineers in taking corrective actions.
Deployment Effectiveness Evaluation.
 The SCELM system has been running stably for over 11 months, monitoring thousands of changes per week. Feedback from Operations and Cloud Engineers (OCEs) indicates that SCELM is capable of detecting nearly all erroneous changes, achieving over 95\% F1-score on classification tasks and around 75\% accuracy in fault localization—results that are highly consistent with our prior experimental evaluations. Compared to traditional manual analysis, SCELM has reduced the time to resolution for erroneous changes by 90\%, significantly improving incident response efficiency. OCEs widely report that SCELM has simplified, accelerated, and automated the change evaluation process. These outcomes highlight SCELM’s strong generalization capability and lay a solid foundation for its broader adoption in future applications.

\section{Discussion}
This study presents SCELM, an intelligent change analysis system designed to unify multimodal data processing and provide interpretable reasoning. Throughout its design, implementation, and long-term deployment, we have gained valuable insights and identified potential threats to the system’s validity.

\subsection{Lessons Learned}
From the development and real-world deployment of SCELM, we have drawn several key lessons:

\textbf{First}, data scarcity, especially during the cold-start phase, emerged as the primary challenge affecting system performance. While our core motivation was to create a unified representation of multimodal data to support downstream tasks, high-quality, fully annotated historical data are often scarce in practical settings. This can lead to suboptimal analysis accuracy in the system's early stages. To address this, practitioners should proactively leverage existing change data to help the model build foundational knowledge. Additionally, systematically organizing internal documentation and historical cases can enrich the context for the RAG-based vector knowledge base, thereby enhancing inference accuracy during online reasoning.

Although this study focuses primarily on RAG due to data constraints, we recognize that fine-tuning large language models (LLMs) could further enhance performance in domain-specific scenarios. Future work will explore the combination of fine-tuning with RAG to optimize performance across diverse settings. Moreover, transfer learning and data augmentation techniques can mitigate cold-start issues by transferring knowledge from similar domains or generating synthetic data to accelerate model adaptation to new environments.

\textbf{Second}, the quality of change ticket information is critical to system performance. We observed that different organizations vary in how they document incidents, and engineers may sometimes fail to report the true root cause accurately. This directly affects both the quality of the unified information carrier and the knowledge base used by RAG. To improve historical data reliability, organizations should ensure that change ticket records are accurate and complete. This is essential not only for retrospective analysis but also for guiding future change evaluations—and is foundational for SCELM’s ability to generate high-quality, interpretable change analysis reports.

\subsection{Threats to Validity}
The primary threats to validity in this study stem from the inherent variability in the capabilities of large models. While SCELM leverages LLMs for expert tasks such as CoT-based reasoning and change analysis generation, its performance can fluctuate due to domain-specific characteristics, the complexity of anomaly detection, and the instability of the models themselves. Even under the same SCELM framework, differences in training data, parameter configurations, and model architectures may lead to varying outcomes. This directly challenges our vision of building a transparent, reliable, and optimizable reasoning engine.

To mitigate these threats, we propose the following strategies:
\begin{itemize}
    \item Ongoing performance evaluation and optimization of multimodal integration strategies. It is essential to select the most appropriate LLM based on the application context and to consider ensemble methods that combine outputs from multiple models to improve overall accuracy. This approach ensures robustness even if the base LLM exhibits performance variability.
    \item Building adaptive evaluation systems that dynamically adjust model weights and strategies based on real-world deployment feedback. For example, the KTO module exemplifies this adaptive optimization by enhancing continual learning from misdiagnosed cases.
\end{itemize}
These strategies will strengthen SCELM’s practical validity and robustness, enabling it to adapt to evolving change scenarios and provide reliable support for decision-making.

\section{Conclusion}
Ensuring reliability during software changes is critical. This paper presents SCELM, an innovative unsupervised framework for automated change evaluation that integrates ECD, FT, and RCCA using multimodal techniques. SCELM leverages large language models (LLMs) and their multitask learning capabilities to fully automate key steps in software change assessment.

While LLMs have been widely applied across domains, our study reveals that their interpretability and adaptability remain limited in the context of change management. The introduction of chain-of-thought (CoT) reasoning significantly enhances the feasibility of deploying LLMs in this domain. Our approach achieves promising results across multiple change datasets.

This work sets a benchmark for future research, establishes a standard for applying LLMs to change management, and contributes to the evaluation methodology for large language models in practical, real-world scenarios.

\bibliographystyle{ACM-Reference-Format}
\bibliography{sample-base-scelm}

\end{document}